\newcommand\lsim{\lesssim}
\newcommand\gsim{\gtrsim}
\newcommand\kb{k_{\rm B}}

\newcommand\np{n_{\rm p}}
\newcommand\nne{n_{\rm e}}
\newcommand\nx{n_{\rm x}}
\newcommand\nt{n_{\rm t}}
\newcommand\hx{H_{\rm x}}

\newcommand\vp{V_{\rm p}}
\newcommand\vx{V_{\rm x}}
\newcommand\tgh{t_{_{\rm gH}}}
\newcommand\tc{t_{_{\rm c}}}
\newcommand\txh{t_{_{\rm xH}}}
\newcommand\dd{{\rm d}}
\newcommand\fgas{f_{_{\rm gas}}}
\newcommand\sxp{\sigma_{_{\rm xp}}}

\documentclass[10pt,preprint2]{aastex}

\begin{document}

\title{Consequences of short range interactions between   dark matter and protons in galaxy clusters}
\author{Leonid Chuzhoy$^{1}$ and Adi Nusser$^{2}$}
\bigskip

\affil{$\ ^1$McDonald Observatory and Department of Astronomy, The University of Texas at Austin, RLM 16.206, Austin, TX 78712, USA; chuzhoy@astro.as.utexas.edu}

\affil{$\ ^2$The Physics Department, The Technion-Israel Institute of
Technology, Technion City, Haifa 32000, Israel; adi@physics.technion.ac.il}

\begin{abstract}

Protons gain energy in short range collisions with 
heavier dark matter particles (DMPs) of comparable velocity dispersion. 
We examine the conditions under which the heating of baryons by scattering off  DMPs  
can  offset radiative cooling in the cores of galaxy clusters. 
Collisions with a constant cross section, $\sxp$,  independent of the relative velocity of
the colliding particles, cannot produce stable thermal
balance. In this case,  avoiding an unrealistic increase of the central temperatures 
yields the upper bound $\sxp< 10^{-25}\;{\rm cm}^2 (m_{\rm x}/m_{\rm p})$, where $m_{\rm x}$ and $m_{\rm p}$ 
 are the DMP and proton mass, respectively. 
A stable  balance, however, can be achieved  for  a power law 
dependence on the relative velocity, $V$, of the form,  $\sxp \propto V^a$ with $a < -3$. An advantage of this heating mechanism is that 
it preserves the metal gradients observed in clusters.

\end{abstract}


\section{Introduction}
\label{int}

While the existence of ``dark'' matter particles (DMPs) is one of the main pillars of the standard cosmological paradigm, the nature of these particles and the way they interact with the ordinary matter are still unknown. Presently the weakly interacting massive particle (WIMP) is the most popular candidate for the role of the DMP, but as long as it remains undetected other candidates present viable alternatives. 
Strongly interacting
heavy particles are produced non-thermally in some cosmological scenarios (e.g. Griest \& Kamionkowski 1990;  Chung, Kolb \& Riotto 1998).   
 Underground experiments, which are primarily designed to search for WIMPs, are insensitive to particles whose interaction cross-section with baryons, $\sxp$, is above $\sim 10^{-30}\;{\rm cm^2}$. Since the flux of such particles is heavily diluted by the ground or even by the atmosphere (Albuquerque \& Baudis 2003), 
the most rigorous  constraints are based  on satellite experiments  (Wandelt et al. 2000) and  observations of astrophysical systems (Starkman et al. 1990; Chen et al. 2002) 
Current constraints allow for super-heavy particles with masses above
$10^{5}\; \rm Gev$ with a wide range of cross sections with ordinary matter.
On the low side, DMPs with $m_{\rm x}<0.4 \; \rm Gev$ are still 
admissible.

In this paper we examine the role
of short range interactions between the DMPs  and protons  in galaxy clusters. 
As particles fall into the  halos,  the gravitational virial theorem implies that baryons and DMPs
acquire similar velocities, $V$. Since the kinetic energy of
a particle of mass $m$  is $m V^2/2$, the direction  of collisional energy
transfer between baryons and DMPs depends mostly on
their masses. For $m_{\rm x}>m_{\rm p}$ and large enough collisional
cross section, heating by DM
 overcomes radiative cooling and increases the temperature.
If this were the case for galaxy clusters, then we would expect to find the largest temperatures at their centers, where 
the density and collision rates are highest.
However, in most  clusters the observed temperature decreases towards the center.
This allows us to place a new upper limit on $\sxp$. For velocity independent scattering the new limit is
two orders of  magnitude lower than the previous limit obtained from
stability analysis of galactic disks and  DM halos \citep{Star}.

Qin \& Wu (2001)  have argued that for $m_{\rm x}\gg m_{\rm p}$ and $\sigma_{\rm xp}\sim
10^{-25} (m_{\rm x}/m_{\rm p})\;{\rm cm^2}$, which is similar to the range of strong interaction,  energy transfer from DM to baryons can explain the deficit of cold gas  ($T<10^7$ K) in the cores of clusters of galaxies, as
suggested by Chandra and XMM-Newton observations  (Kaastra 2001, Peterson 2001, Tamura 2001). 
Here we elaborate on the work of 
Qin and Wu (2001) in several ways. 
We use a more accurate description of the  radiative cooling and heating rates and  consider the
consequences of a velocity dependent cross section. Further, we account for the 
increase in the velocity dispersion of the dark matter as it contracts while 
transferring energy into the protons.
We show that 
 unless the cross-section strongly varies with energy,  a state of thermal equilibrium between heating by the DMPs and radiative cooling is unstable.

The paper  is organized as follows.  In \S 2 we write down the 
relevant equations. In  \S 3 we check the requirements for the stable thermal balance between radiative cooling and heating by DMPs.
 In \S 4 we derive analytic upper limits on the cross sections.
We present numerical solutions in \S 5 and conclude with a  
discussion in \S 6.

\section{The equations}

We begin by writing down the equations governing the temperature and density of  a  gas element subject to radiative cooling and heating by collisions with dark matter particles.
As the self-gravitating  dark matter particles transfer energy to the baryons, they 
get closer to the center and become hotter. Although the mass dark matter 
particle is  much larger than that of the proton, the energy density
is not and therefore we must account for the increase in the 
velocity dispersion of the dark matter particles. We denote the gas heating and cooling rates, both per unit volume, by $\hx$ and $L$, respectively.  
The corresponding rates per unit mass are
$\hx/n_{\rm t} $ and $L/n_{\rm t}$ where 
$n_{\rm t}\approx 2.2 \np$ is the total number density of 
all particles in a fully ionized primordial 
plasma. 
The energy balance  equation is most conveniently expressed 
 in terms of 
the entropy. 
We 
define an  "entropy" as $S=k_{\rm B} T/{\nne}^{2/3}$ where $\nne\approx 1.16\np $  is the electron number density in $\rm cm^{-3}$ and $\kb T$ is in $\rm kev$. 
The entropy obeys, 
\begin{equation}
\frac{\dd \ln S}{\dd t}=\frac{2}{3\nt \kb T}\left(\hx-L\right) \; .
\label{eq:ent}
\end{equation}
The advantage of this form of the energy balance equation is that the adiabatic heating/cooling 
is readily included. 

As will be shown below, in the first approximation the heating rate $\hx$ depends on the dark matter properties
only through the energy density (per unit volume) 
$E_{\rm x}=(1/2) n_{\rm x}m_{\rm x} V_{\rm x}^2$ where $n_{\rm x}$ is the number density of dark matter particles, and
$V_{\rm x}$ is their velocity dispersion. 
According to the  virial theorem $E_{\rm x}=-E_{\rm total}$. Therefore,  
\begin{equation}
\label{eq:virial}
\frac{\dd E_{\rm x}}{\dd t}=H_{\rm x}\; .
\end{equation}
There are three time-scales in the problem, $\tc$, $\tgh$, and $\txh$ which 
correspond, respectively, to  
the radiative cooling, heating of  gas by collisions with DM particles, $\tgh$,  
and  kinetic energy gain of dark matter. These time scale are given by
\begin{eqnarray}
\tc&=& \frac{3\kb \nt T}{2 L}\\
\tgh&=&\frac{3\kb \nt T}{2 \hx}\\
\txh&=& \frac{E_{\rm x}}{\hx}= \frac{E_{\rm x}}{E_{\rm gas}}   \tgh.
\end{eqnarray}
The virial theorem implies that after the collapse the energy densities of DM and gas relate as their mass densities, $E_{\rm gas}/E_{\rm x}\approx\fgas\approx 0.15$. 
However, the energy acquired by protons during the collapse is very quickly 
shared  with electrons and helium ions so that the velocity dispersion of 
these species is fixed by energy equipartition. Therefore, initially we have $n_{\rm p}m_{\rm p}\vp^2/n_{\rm x}m_{\rm x}\vx^2=(n_{\rm p}/n_{\rm t})\fgas\approx 0.07$.
 Subsequent effects of DM-baryon interaction 
 are then determined by the equations (\ref{eq:ent}) and (\ref{eq:virial}). 
The cross section $\sxp$ must satisfy the condition 
\begin{equation}
\tc<\tgh\ \; ,
\label{eq:lim2}
\end{equation} 
otherwise the temperature would increase at the 
central regions, contrary to observations. 
 The last condition is sustained by  hydrodynamical simulations 
which show  that as clusters form their temperature profiles increase towards the center (e.g. Navarro, Frenk \& White 1995). Therefore,  if the heating by DMPs is stronger than cooling then  the temperature will always be highest at the center.

If the mass and typical kinetic energy of a DMP are much larger than 
 those of a proton,   $m_{\rm x}\gg m_{\rm p}$ and  $m_{\rm x}V_x^2\gg  m_{\rm p}V_p^2$, then the DMP recoil in a single scattering with proton can be neglected. In this case,
the gas heating rate per unit volume is (Qin \& Wu 2001) 
\begin{eqnarray}
\label{QW}
H_{\rm x}=\frac{m_{\rm p}n_{\rm p} n_{\rm x}\sigma_{\rm xp}}{2\pi V_{\rm x}^3V_{\rm p}^3}
\int_0^{\infty} u^2v^2e^{-u^2/2V_{\rm p}^2-v^2/2V_{\rm x}^2}dvdu \nonumber \\
\times \int^{1}_{-1}d\cos\theta(s^2+v^2-u^2)s,
\end{eqnarray}
where we have assumed Maxwellian energy distribution for 
the DMPs and protons, $V_{\rm x}$ and $V_{\rm p}$ are, respectively,  the velocity dispersions of DMPs and protons, and $\cos\theta=(u^2+v^2-s^2)/2uv$.
 Assuming a power law dependence , $\sigma_{\rm xp}=\sigma_0 (s/V_0)^a$, of the cross-section on 
the relative speed $s=|\vx-\vp|$ transforms  eq. \ref{QW} into
\begin{eqnarray}
\label{QW2}
H_{\rm x}=\frac{m_{\rm p}n_{\rm p} n_{\rm x}\sigma_0 V_0^{-a}}{2\pi V_{\rm x}^3V_{\rm p}^3}
\int_0^{\infty} u^2v^2e^{-u^2/2V_{\rm p}^2-v^2/2V_{\rm x}^2}dvdu \nonumber \\
\times \int^{1}_{-1}d\cos\theta(s^2+v^2-u^2)s^{1+a}.
\end{eqnarray}
For integer values of $a$ the above integrals can be solved analytically. However, the solutions are quite cumbersome, so it is more convenient to use an approximation
\begin{eqnarray}
\label{h0}
H_{\rm x}\approx 6\cdot 2^{a}m_{\rm p}n_{\rm p} n_{\rm x}\sigma_0   V_0^{-a}V_{\rm x}^{3+a}(1+\frac{V_{\rm p}^2}{V_{\rm x}^2})^{\frac{1+a}{2}},
\end{eqnarray}
In the central regions of clusters $V_{\rm x} \lsim V_{\rm p}$, which allows further simplification of eq. \ref{h0}
\begin{eqnarray}
\label{h01}
H_{\rm x}\approx 6\cdot 2^{a} m_{\rm p}n_{\rm p} n_{\rm x}\sigma_0  V_0^{-a}V_{\rm x}^2 V_{\rm p},
\end{eqnarray}


We write the cooling rate as (Tozzi \& Norman 2001)
as 
\begin{equation}
L=n_{\rm i}n_{\rm e}\left[ C_1 (\kb T)^\alpha +C_2(\kb T)^\beta +C_3 \right] \; 
\label{eq:tozzi} 
\end{equation}
in units of  $10^{-22}\rm erg
\; cm^{-3}\;  s^{-1} $. The number density of ions $n_{\rm i}\approx 1.08\np $  and electrons $\nne\approx 1.16\np $ 
 are in $\rm cm^{-3}$. For metallicity  of one third solar, $\alpha=-1.7$,
$\beta= 0.5 $, $C_1= 8.6\times 10^{-3} $, $C_2= 5.8\times 10^{-2} $, and $C_3=6.4\times 10^{-2}$ and $\kb T$ is in kev. 

\section{Thermal stability analysis}

Because the term $\dd E_{\rm x}/\dd t$ never vanishes (see equation 
\ref{eq:virial}), a strict thermal equilibrium 
cannot hold. However, the time scale, $\txh$, for   significant changes in $E_{\rm x}$  is typically a factor of $1/\fgas\approx 6$  
longer than the time scale, $\tgh$, for heating the baryons.  
Therefore, it is meaningful  to address the question of whether  
a quasi-equilibrium with $\dd S/\dd t\approx 0$ (see equation \ref{eq:ent}) is thermally stable over time-scales shorter than  $\txh$.
We express $\hx$ and $L$  in terms of $T$ and the 
pressure $p=\nt \kb T$. We take $\hx=A p T^{a/2-1}$ 
and  $L=B T^{\beta-2} p^2$
where $A$ and $B$ represents all quantities that 
are independent of $p$ and $T$.
 Further, we write 
 $\ln S=\ln T^{5/3}/p +const$. 
Equation (\ref{eq:ent}) is then   
\begin{equation}
\frac{3}{2}p\frac{\dd \ln (T^{5/3}/p)}{\dd t}= A p T^{a/2-1/2}-
B p^2T^{\beta-2} \; . 
\end{equation}
Let the r.h.s vanish for $T={T}_0$ at $p=p_0$
and consider variations in $T$ at constant pressure $p=p_0$. 
Substituting $T=T_0 +\delta T$,  the last equation gives
$ (5/2)p\dot T=(a/2+3/2-\beta)ApT^{a/2-1/2}$ to first order in
 $\delta T$. 
Therefore, the system is stable for  $a <2\beta-3$.
For $T\approx 10^8\rm K$ and metallicity of one third solar, $\beta\approx 0.5$, so that $\alpha<-2$. For clusters with lower temperatures
and higher metallicity $\beta\approx -0.3$, and we get $\alpha<-3$.
Therefore, the dependence of the cross section must be steep
in order to obtain thermal stability over 
time scales $<\txh$. 

\section{Upper limits on cross sections}
\label{upper}
To allow the temperature at the core to fall the cooling must overcome the heating, $L>H$. Since both $L$ and $H$ are not constant, the value of $\sxp$ which gives the thermal balance changes over time. However, unless the condition $L>H$ is not satisfied from the start, it will never be satisfied later. Thus we may derive the upper limit on the cross-section by using the initial conditions of clusters. 

For $a\gsim-2.3$ the stringiest limit is obtained from the hottest clusters where the cooling is dominated by bremsstrahlung and the cooling function is well approximated by
\begin{eqnarray}
\label{L1}
L=2.9\times 10^{-27}n_p^2 T^{0.5}\; {\rm erg\; s^{-1}\;  cm^{3}} \; ,
\label{eq:lapp}
\end{eqnarray}
where $T$ is in K. 
The limit on $\sxp$ is obtained from the condition
(\ref{eq:lim2}) 
using the expressions (\ref{h01}) and (\ref{eq:lapp}). 
Taking $n_{\rm p}m_{\rm p}\vp^2/n_{\rm x}m_{\rm x}\vx^2\approx 0.07$,
the condition (\ref{eq:lim2}) yields\footnote{The upper limits on the
cross sections do not depend on the actual value of $\np$ since
for a given $\fgas$ both the initial cooling and heating rate 
are proportional to $\np^2$.}
\begin{equation}
\label{lim2}
\sxp< 2.7\times 10^{-25}{\rm cm^2}\;  \cdot 2^{-a}
\left(\frac{m_{\rm x}}{m_{\rm p}}\right)\left(\frac{T}{10^8\; {\rm K}}\right)^{-1}\left(\frac{V_0}{1\;{\rm m/s}}\right)^{a}.
\end{equation}
For energy independent scattering  ($a=0$) we get 
\begin{equation}
\label{lim3a}
\sxp< 2.7\times 10^{-25}  {\rm cm^2}\;  \left(\frac{T}{10^8\; K}\right)^{-1}\left(\frac{m_{\rm x}}{m_{\rm p}}\right)\;,
\end{equation}
which for the hottest clusters, such as  RX J1347.5-1145 with $T\gsim 1.7\cdot 10^8$ K (Allen, Scmidt \& Fabian 2002), yields
\begin{equation}
\label{lim3}
\sxp<  1.5 \times 10^{-25}  {\rm cm^2} \left(\frac{m_{\rm x}}{m_{\rm p}}\right)\; .
\end{equation}
For $a\lsim-2.3$ it is the coldest clusters ($T\sim 3 \cdot 10^8$ K) which give the stringiest limit. There the eq. \ref{L1} is no longer a very good approximation and the 
full cooling function has to be used. Therefore we present the limit graphically in  fig. \ref{figX}.

The above calculations ignore several additional heating processes such as AGN, thermal conduction and SN feedback. However, if these are important they will make our limits even more restrictive.

\section{Numerical solutions}
We show now results of  numerical integration
of the equations (\ref{eq:ent}) and (\ref{eq:virial}),
for various values of initial temperatures $T$ using the 
 full expression for the 
cooling rate given by (\ref{eq:tozzi}). 
Our goal is to confirm the upper limit derived in the previous section 
and to demonstrate that a reasonable temperature evolution is 
obtained for energy dependent cross sections. 

We would like to integrate the equations 
with initial conditions describing the physical state in the cool 
cores of  clusters 
before cooling and heating become important.
The initial temperature in the core can be extrapolated from the outer regions 
assuming an initial isothermal profile. We also need 
the initial number density, $\np$,  in the core. 
This is more tricky to obtain as the observed 
$\np$ is likely to be substantially different 
from its initial value and from the 
measured value in the outer regions. 
To illustrate the effect of varying $\np$ we experiment here with 
two choices for $\np$.
We use  three values for the initial
temperatures, $T=16$, 8 and 3 kev.
Candidates for clusters with these temperatures are, respectively,   
RX J1347.5-1145 (Allen, Schmidt \& Fabian 2002 ,  A2204 (Pointecouteau, 
Arnaud \& Pratt 2005), 
and A1991 (Pratt \& Arnaud 2005)
respectively.  
For each temperature, we integrate the equations for 
two values of the initial number density: $\np=0.015$ and 0.007 $\rm cm^{-3}$. 
At the initial time  the condition 
$n_{\rm p}m_{\rm p}\vp^2/n_{\rm x}m_{\rm x}\vx^2=(n_{\rm p}/n_{\rm t})\fgas$ is assumed to hold.  
Note that for a NFW density profile (Navarro, Frenk \& White 1997) 
the product $\nx \vx^2$ is independent of position in the central regions. 
We further  assume that the gas pressure ($\propto n_{\rm t} T$) is 
equal to its initial value.  
The equations are integrated forward for various values of $T$.

\begin{figure}
\epsscale{1.0}
\plotone{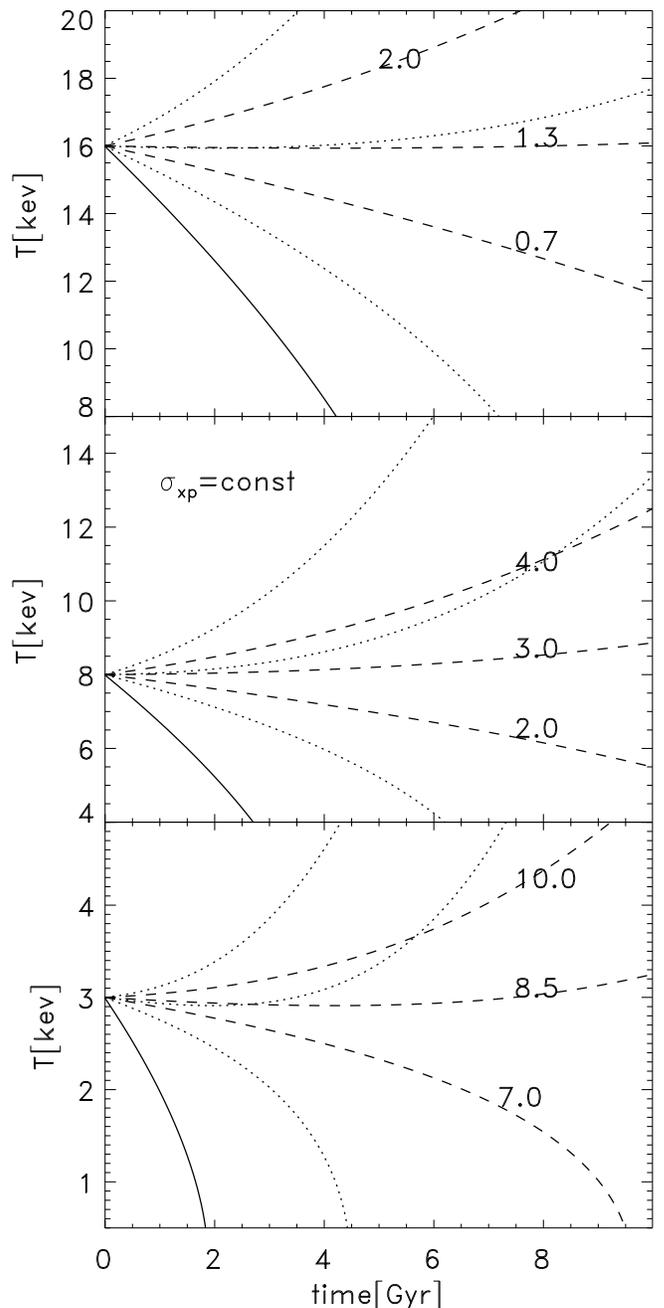}
\caption{The  temperature evolution of a gas element subject to 
radiative cooling and heating by collisions with dark matter particles
 for 
constant $\sxp$ ($a=0$) and $\vx/\vp=0.5$. 
The top, middle and bottom panels correspond  to 
 initial temperatures of  16, 8 and $3\rm \; kev$, respectively. 
The solid curves show are obtained obtained with
radiative cooling only. 
Dashed and dotted curves are, respectively,  for  $\np=0.007$ and 
$0.015\; \rm cm^{-3}$, at the initial time.
The labels on the dashed  curve indicate $\sigma_{\rm xp}$ in units 
of $10^{-25}(m_{\rm x}/m_{\rm p}){\rm cm}^2$. 
}
\label{fig1}
\end{figure}

\begin{figure}
\epsscale{1.0}
\plotone{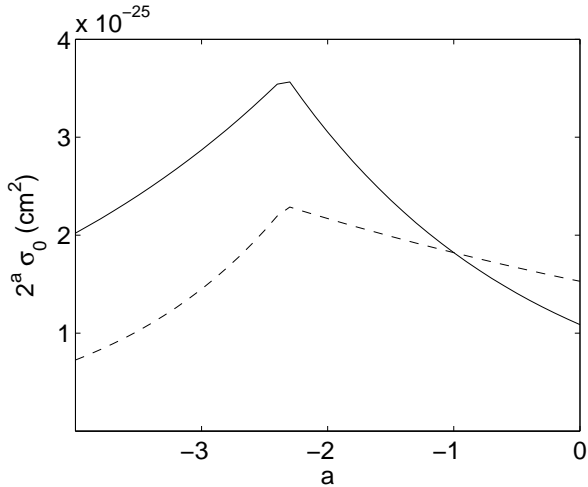}
\caption{The upper limit on $\sigma_0$ for $V_0=10^3$ km/s. The solid and the dashed curves correspond to $\vx/\vp=1$ and 0.1 respectively.
}
\label{figX}
\end{figure}

Figure (\ref{fig1}) shows the temperature evolution for 
for  velocity independent
cross sections, i.e. $a=0$.
The three panels show the temperature evolution
for three respective values of the initial temperature
and initial $\np$ of $0.007$ (dashed curves) and
$0.015\; \rm cm^{-3}$ (dotted). 
The solid curve  in each panel label is obtained 
with radiative cooling only. 
In  each panel  three
different values of $\sxp$ have been used. 
These values, in units of $10^{-25}m_{\rm x}/m_{\rm p} \; \rm cm^2$, 
are  indicated
by the numbers attached to the dashed curves.
These three values are also used in obtaining 
the dotted  curves (higher values  correspond to higher 
curves).
No single value of the cross section is able to balance 
 radiative  cooling  for all temperatures.
 The value of $\np$ clearly affects the time-scale 
 of the evolution, with the larger $\np$ giving the faster 
 evolution. 
An inspection of this figure allows us to place an upper 
limit on  
the cross section. The strongest constraint is derived 
from the highest temperature observed clusters
of $T\sim 15\rm kev$ (Allen, Schmidt \& Fabian 2002).
According to  top panel  $\sxp<10^{-25}m_{\rm x}/m_{\rm p} \rm \; cm^2$, 
consistent with our previous 
estimate in \S\ref{upper}. Exceeding this limit
produces temperature profiles rising towards the center of all 
clusters, contrary to
observations.  The limit is insensitive to $\np$.

\begin{figure}
\epsscale{1.}
\plotone{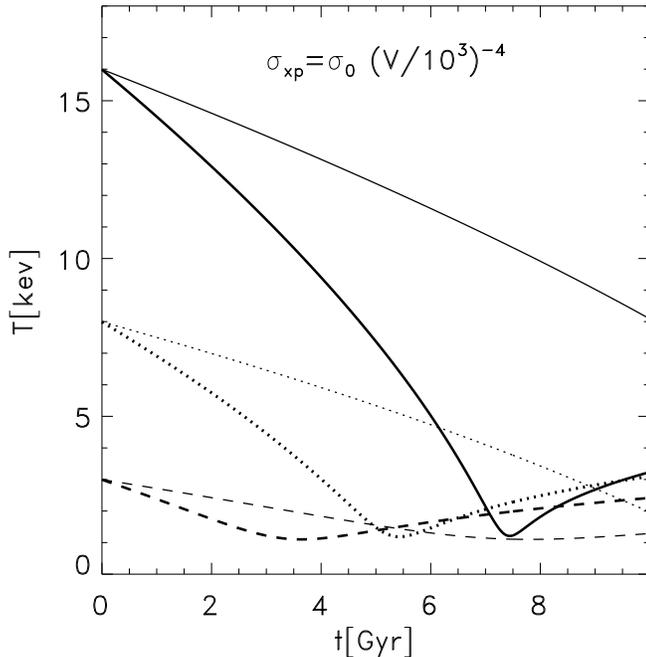}
\caption{The temperature evolution 
assuming $\sxp=\sigma_0(V/V_0)^{-4}$ 
The parameter $\sigma_0$ is given by $\sigma_0=10^{-25}m_{\rm x}/m_{\rm p}\; \rm cm^{-2}$ and $V_0=10^3\rm \; km/s$. The 
solid, dotted and dashed lines correspond to initial
temperatures of $16$, 8 and 3 kev, respectively. 
Thick and thin lines are for  initial $\np$ of $0.015$ and 
$0.007\rm \; cm^{-3}$, respectively. For reference, the look-back time to z=1 is 8 Gyrs.}
\label{fig2}
\end{figure}

In figure  (\ref{fig2}), we show the temperature 
evolution  obtained from the clusters obtained with 
 $\sxp=\sigma_0 (V/V_0)^{-4}$ where $\sigma_0=
  10^{-25}(m_{\rm x}/m_{\rm p})\; \rm cm^2$ and $V_0=10^3\rm km/s$.
  The thin and thick lines correspond to $\np=0.007$ and 
  0.015 $\rm cm^{-3}$, respectively. The solid,  dotted and
  dashed are for initial temperatures of $16$, 8 and $3\rm \; kev$.  
The late time temperature increase seen for some of the curves 
is caused by the enhancement of heating as a result of 
virial  contraction of the dark matter halo. 
However, even for initial $\np=0.015\; \rm cm^{-3}$,
 the temperature increase is minor  at $t=8\rm \; Gyr$, corresponding to
a redshift of $z=1$. 
Overall,  we conclude a cross section with a steep dependence on energy 
can balance  cooling  in clusters of galaxies.

\section{Discussion}

\begin{figure}
\epsscale{1.}
\plotone{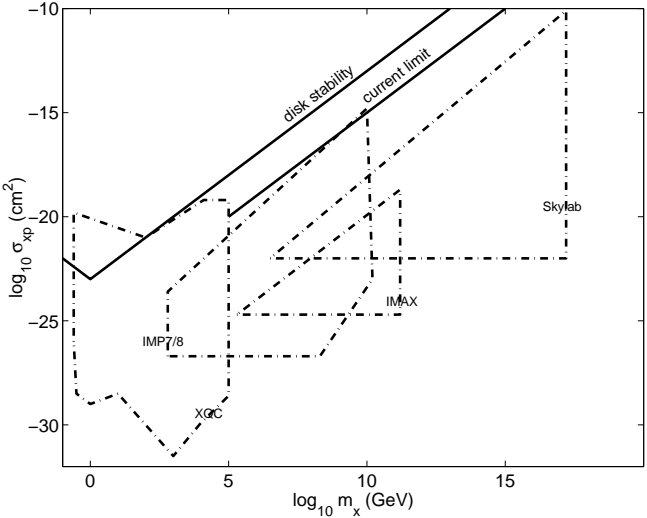}
\caption{
A summary of of the constraints on the 
cross section. Our upper limit is derived 
for cross sections that are independent of 
energy. See Wandelt et al.(2000) for details. }
\label{fig4}
\end{figure}

Qin \& Wu (2001) have suggested that collisional  energy transfer from  
heavy dark matter  particles ($m_{\rm x}/m_{\rm p}>10^5$) 
to protons can balance cooling and thus explain the deficit of the cold gas in clusters (Kaastra 2001, Peterson 2001, Tamura 2001).
In the original Qin \& Wu scenario the cross section is 
assumed to be a constant independent of energy (i.e. the relative 
velocity between the interacting proton and DM particle).
We have found that $a)$  heating of DM with a constant cross section cannot offset cooling for clusters of different virial temperatures, and $b)$
 any thermal equilibrium obtained with a  constant $\sigma_{\rm xp}$, is unstable. 
 However, an significant upper limit on  cross sections
 that are independent of velocity 
 can  be obtained by demanding the temperature not to rise towards the center, as seen in observations. Our  upper limit on $\sxp$ versus the mass is represented  in the diagram shown in figure  (\ref{fig4}) which also shows previous limits
 (Wandelt et al. 2000).  
 
Our study incorporates the contraction of dark matter halo as it 
transfers energy to the protons. This contractions leads to an increase
in the velocity dispersion of DM particles and hence to an enhancement 
of the heating rate. However, for time scales short compared with $\txh$ one can study the thermal stability of the system.
We have found that for $\sxp=\sigma_0(V/V_0)^a$ a stable thermal equilibrium  is possible for $a<-3$ (for  $T>2$ keV and 
metallicity below 0.6 solar). 
For strong interactions the  
scattering cross-section is expected to have a weak dependence on velocity. However, electromagnetic interactions with $a=-4$ may fit very well. The DMP with a unit charge and mass a few times above $10^5$ GeV, which is just outside the conservative limit on charged DMP (De Rujula et al. 1990), seems optimal for suppressing the cooling flow.
Furthermore DMPs of such mass if produced in thermal equilibrium would explain the present value of $\Omega_{\rm m}$ (Griest \& Kamionkowski 1990).
Another attractive feature of 
heating by dark matter  is that it   
preserves the metal gradient seen in observations of some clusters
(e.g. Bohringer et al. 2004).
 This is advantageous 
over mechanical  sources of heating such
as AGN feedback  which tend to reduce the metal gradients.

An alternative possibility for stable thermal balance is a DMP with mass comparable to $m_{\rm p}$. In this case protons and DMPs have comparable energies.  
Thus no matter how large  $\sxp$ is, no runaway heating will occur. One problem with this scenario is that it leaves a very narrow range for $m_{\rm x}$. 
For $m_{\rm x}>0.4$ GeV the allowed values of $\sxp$ are too low to achieve thermal balance (Wandelt et al. 2000).
On the other hand DMPs with  $m_{\rm x}\lsim 0.2-0.3$ GeV will typically be less energetic than baryons at the central region. A second problem is that observationally the baryon fraction declines towards the center. If the DMPs and protons are in thermal balance, then the DMPs  with $m_{\rm x}<\mu m_{\rm p}$ can have steeper density profile only if their velocity distribution has significant anisotropy.

The calculations we have done  for protons can be extended to other gas particles, such as electrons and helium ions. Thus for velocity independent interaction we find that for helium $\sigma_{\rm xHe}<1.2\cdot 10^{-24}(m_{\rm x}/m_{\rm p})\;{\rm cm^2}$, while for electrons with  $V_{\rm e}=\vp(m_{\rm p}/m_{\rm e})^{1/2}$ and  $\nne=1.16\np$, the limit is
$\sigma_{\rm xe}<6\cdot 10^{-24}(m_{\rm x}/m_{\rm p})\;{\rm cm^2}$.

Our analytic treatment is based on the local values of physical parameters and therefore can not accurately account for the potentially important processes of convection or 
the change of central pressure in response to the changing temperatures. To describe these effects in detail 
a hydrodynamical simulation is needed. This is feasible since the relevant heating function is easy to incorporate in these simulations.

\acknowledgments

AN thanks Jerry Ostriker, Joseph Silk and 
Etienne Pointecouteau for stimulating discussions.
He also thanks the National Astronomical observatory of Japan and 
the Oxford Astrophysics Department for the hospitality and support.
We thank Paul Steinherdt for allowing the
use of his figure.

\end{document}